\documentclass{aastex}

\slugcomment{(Revision 23.2, 06 Oct 2000)}

\shorttitle{Schneider et al.}
\shortauthors{55~Cancri}

\begin{document}

\title{NICMOS Coronagraphic Observations of 55~Cancri}

\author{G. Schneider}
\affil{Steward Observatory, The University of Arizona, 
Tucson, AZ 85721}
\email{gschneider@as.arizona.edu}

\author{E.E. Becklin}
\affil{Dept. of Physics and Astronomy, University of 
California, Los Angeles, Los Angeles, CA 90095}

\author{B.A. Smith}
\affil{Institute for Astronomy, University of Hawaii, 
Honolulu, HI 96822}

\author{A.J. Weinberger}
\affil{Dept. of Physics and Astronomy, University of 
California, Los Angeles, Los Angeles, CA 90095}

\and

\author{M. Silverstone and D.C. Hines}
\affil{Steward Observatory, The University of Arizona, 
Tucson, AZ 85721}

\begin{abstract}
We present new near-infrared (1.1-$\mu$m) observations of the 
circumstellar environment of the planet-bearing star 55~Cancri.  With 
these Hubble Space Telescope images we are unable to confirm the 
observation of bright scattered radiation at longer NIR wavelengths 
previously reported by \citet{TB} and \citet{TBR2000}.  NICMOS 
coronagraphic images with detection sensitivities to 
$\sim$100~$\mu$Jy~arcsec$^{-2}$ at 1.1~$\mu$m in the region 28 - 60 
AU from the star fail to reveal any significant excess flux in point- 
spread-function (PSF) subtracted images taken in two Hubble Space 
Telescope orbits.  These new observations place flux densities in the 
19-28 AU zone at a factor of ten or more below the reported 
ground-based observations.  Applying a suite of a dozen well-matched 
coronagraphic reference PSFs, including one obtained in the same 
orbits as the observations of 55~Cnc, yielded consistently null 
results in detecting a disk.  We also searched for, and failed to 
find, a suggested flux-excess anisotropy in the ratio of $\sim$1.7:1 
in the circumstellar background along and orthogonal to the plane of 
the putative disk.  We suggest that, if such a disk does exist, then 
the total 1.1-$\mu$m spectral flux density in an annular zone 28 - 42 
AU from the star must be no more than $\sim$~0.4mJy, at least ten times 
smaller than suggested by Trilling and Brown, upon which their very 
large estimate for the total dust mass (0.4~$M_{\earth}$) was based. 
Based on the far infrared and submillimeter flux of this system and 
observations of scattered light and thermal emission from other debris 
disks, we also expect the intensity of the scattered light to be at 
least an order of magnitude below our upper limits.
\end{abstract}


\section{Introduction}

The connection between the formation and growth of planetesimals into 
larger bodies and the circumstellar debris disks around young and 
nascent stars from which they arise has been the subject of continual 
speculation, discussion and inquiry, since the original scenario for 
such evolution was first expanded upon by \citet{Laplace} following 
the general proposition by \citet{Kant}.  With the imaging of the 
$\beta$ Pictoris circumstellar disk by \citet{Smith}, as originally 
detected by IRAS \citep{Aumann}, this paradigm of the formation of 
planetary systems became more visceral.  The observation and direct 
imaging of such circumstellar disks, seen in scattered light 
surrounding other stars and stars exhibiting morphologically resolved 
spatial features in the near infrared (NIR), has recently been 
reported (e.g., HR~4796A; \citep{Schneider99}, HD~141569, 
\citep{Weinberger99}; TW~Hydrae, \citep{Weinberger00}).  The 
expectation for imaging dust disks around such stars was hopeful, 
given their relatively large thermal infrared excesses.  The detection 
and imaging of possible circumstellar material around stars having 
little or no thermal infrared or submillimeter radiation was presumed 
to be much more difficult.  Nonetheless, a survey of solar-like, 
planet-bearing stars as ascertained from radial velocity surveys was 
conducted by \citet{TB} [henceforth TB98] who reported on the 
detection of a bright dust disk seen in scattered light around 
55~Cancri.

55~Cancri ($\rho^{1}$ Cnc, HR~3522, HD~75732, G8V, d = 12.53 pc., V = 
5.93 \citep{Marlborough}, H = 4.17 \citep{Persson}, [RA=08H 52M 36.13S 
, DEC=+28\degr\ 19' 53\farcs0, J2000.0, epoch 1991.25]) is an older 
star ($\sim$5~Gyr, \citet{Baliunas}) for which a planetary companion 
of Msini = 0.84~$M_{Jup}$ was found by \citet{Butler} in their radial 
velocity search for exosolar planets \citep{Marcy94}.  With a somewhat 
peculiar spectrum, \cite{Cowley} classified it as G8V, and hence 
``solar-like''.  More recently \cite{Baliunas} found its spectrum more 
closely resembled a sub- giant and reclassified it as G8II-III, 
although a search through the literature spanning more than four 
decades (e.g., \citet{Nikonov}, \citet{Marlborough}, and others) 
reveals no evidence of any optical variability as may be expected at 
some level for a sub-giant.  \citet{Gonzalez98a} confirmed its 
sub-giant luminosity reclassification but noted that its 
theoretical-isochrone age is anomalous.  Further they tentatively 
suggest the presence of an unresolved binary companion orbiting nearly 
pole-on, and that their results are consistent with a gas giant planet 
encroaching very close to the star and accreting H and He depleted 
material onto the photosphere.  A period of 14\fd65 was ascribed to the 
planetary companion by \citet{Butler}.  At a distance of 12.53~pc, as 
determined by Hipparcos, the companion orbits with a semimajor axis of 
0.12~AU and eccentricity of 0.03.

TB98 reported the detection of a disk using near-infrared 
coronagraphic imaging at 1.62~$\mu$m, 2.12~$\mu$m, and 2.28~$\mu$m, 
using the Cold Coronagraph (CoCo)/NSFCAM \citep{Toomey} with a 
1\farcs5\ radius mask at the IRTF. The maximum H-band brightness of 
the disk was given as $\sim$1~mJy~arcsec$^{-2}$ at a radius of 
2\arcsec\ .  The disk reported was asymmetric in the sense that the 
total flux density in quadrants described to be along the major axis 
was $\sim$70\% higher than the flux in the minor axis quadrants.  The 
radial decrease in brightness was found to follow a r$^{-5}$ power law 
and the inferred mass in dust was 0.4~$M_{\earth}$.  \citet{TBR2000}, 
(henceforth TBR2000) later suggested that a flux density peak in 
H-band might approach 3~mJy~arcsec$^{- 2}$, as discussed in the 
context of a survey of six stars with known radial velocity 
companions.

Evidence for the detection of a Vega-like disk at 60~$\mu$m with a 
flux excess of 170~$\pm$30~mJy was reported by \citet{Dominik} from 
ISOPHOT observations with no detection at either 25~$\mu$m or longer 
(95~$\mu$m, 160~$\mu$m, 185~$\mu$m) wavelengths, from which they 
inferred a disk radius of $\sim$50~AU. The corresponding $L_{\rm 
disk}/L_{*}$ at 60~$\mu$m is 6x10$^{-5}$, implying two orders of 
magnitude less dust than in other disks seen in scattered NIR light, 
as discussed above.  Recently, \citet{Jayawardhana} (henceforth 
JAY2000) reported a submillimeter excess of 7.5$\pm$4.2~mJy at 
450~$\mu$m and, more significantly, 2.8$\pm$0.5~mJy at 850~$\mu$m, as 
determined from SCUBA/JCMT observations.  They attributed this excess 
to thermal emission from a dust population resembling our Kuiper Belt.  
The amount of dust required to produce these submillimeter fluxes is, 
again, two orders of magnitude below the dust seen in the previously 
noted near-infrared scattered light disks.  Mid-IR (10.8-$\mu$m and 
18.2-$\mu$m) images obtained by JAY2000 using the University of 
Florida's OSCIR camera on Keck II showed no excess emission above the 
stellar photospheric level.

In an attempt to confirm and further characterize the reported NIR 
scattered light disk we have imaged the circumstellar environment of 
55~Cnc at four wavelengths with the coronagraph in the Near Infrared 
Camera and Multi-Object Spectrometer (NICMOS) on the Hubble Space 
Telescope (HST).  Given the sensitivities routinely achieved by NICMOS 
differential coronography, we had anticipated detecting and imaging 
the disk readily at the $\sim$2\arcsec\ limiting radius of the CoCo 
observations based on the flux levels reported by TB98.  From our 
NICMOS observations, however, we are unable to confirm the existence 
of a NIR flux excess in our most sensitive (1.1-$\mu$m) images at flux 
density levels an order of magnitude lower than those reported by 
TB98.

\section{Observation}

Near infrared coronagraphic observations of 55~Cnc were obtained on 1 
November 1998 with four filters (Table~1) in NICMOS camera 2 (image 
scale: X = 0\farcs0760 pixel$^{-1}$, Y = 0\farcs0753 
pixel$^{-1}$) in two Hubble Space Telescope (HST) orbits (see 
Table~2).  The star was occulted in the 0\farcs3 radius 
coronagraphic hole, 3\farcs4 and 5\farcs5 from the (-X, +Y) 
corner of the 19\farcs46 x 19\farcs28 field of view.  After 
acquiring the target into the low-scatter point of the coronagraphic 
system, repeated MULTIACCUM imaging sequences were executed at two 
field orientations (spacecraft roll angles constrained by available 
guide stars) differing by 15\degr.

The imaging strategy for this two-orbit observation differed somewhat 
from that which we routinely employed for other targets in our 
HST/NICMOS disk and companion search programs \citep{Lowrance98}.  
Those single-orbit, single-color, observations were designed as the 
initial detection phase of an unbiased multiple target survey.  Here, 
in this two-orbit observation, we were attempting to confirm and 
further characterize the disk reported by TB98.  At the first field 
orientation a deep-imaging sequence was executed at 1.1~$\mu$m, where 
the residual instrumental scatter in the coronagraph is minimized by 
the small size of the PSF relative to that of the coronagraphic hole 
\citep{Schneider98b}.  At the same orientation, images were also 
obtained at 2.04~$\mu$m and 2.40~$\mu$m to provide additional color 
information on the disk.  In the following orbit the spacecraft was 
rolled by 15\degr\ and additional 1.1-$\mu$m images were obtained to 
discriminate between artifacts induced by the HST+NICMOS optical 
system and true circumstellar features.  In the second orbit we also 
acquired a short 1.6-$\mu$m image to allow us to directly compare the 
morphological and photometric properties of the putative disk in the 
same spectral region as reported by TB98.  In the same orbit the 
spacecraft was then slewed a small distance ($\sim$1\fdg3) to obtain 
images to construct a contemporaneous 1.1-$\mu$m PSF from a reference 
star, HD~75216 (SAO 80447, PPM 99074, K2III, V = 7.38, H = 4.60, 
[RA=08H 49M 45.32S, DEC=+29\degr\ 26' 56\farcs2, J2000.0, epoch 
1991.25]), with a spectral type and brightness similar to 55~Cnc.

\section{Data Reduction}

The individual reads in each raw MULTIACCUM image \citep{Mackenty} 
were dark subtracted, corrected for non-linearities in pixel 
responses, flat-fielded and combined into a count-rate image (counts 
sec$^{-1}$ pixel$^{-1}$) employing an analog to the CALNICA software 
used in the STScI data pipeline.  Calibration reference files (STEP32 
and STEP64 dark frames, and flat fields for the filters employed) 
prepared by the NICMOS Instrument Definition Team (IDT) were used in 
processing the images, rather than library reference files supplied by 
STScI. The reference flat fields were augmented with renormalized data 
from contemporaneous lamp calibration images (obtained as part of the 
target acquisition process) to permit calibration very close to the 
edge of the coronagraphic hole, since its position in the NICMOS focal 
plane is known to vary slightly with time.  Bad pixels were replaced 
in the count-rate images by 2D Gaussian weighted interpolation of 
neighbor pixels with radii appropriate for the wavelengths of the 
different filters.  Each reduced set of images with the same target, 
filter and orientation were then median combined into a single image.  
Known artifacts induced by the readout electronics, such as broad 
``stripes'' orthogonal to the readout direction and localized elevated 
count levels at the locations of deeply exposed targets, both 
replicated in all four detector quadrants, were characterized and 
removed.  Count-rates were converted to physical flux densities based 
on photometric calibrations derived as part of the overall 
instrumental calibration by the NICMOS IDT (Rieke 1999, private 
communication), as shown in Table~1.

\section{PSF Subtracted Images (1.1~$\mu$m)}

We have searched our 55~Cnc images obtained in all four spectral bands 
(Table~1) for evidence of light scattered by circumstellar material.  
In this paper we discuss only our 1.1-$\mu$m observations, which 
afford the greatest sensitivity for disk detection, as the stray light 
rejection of the NICMOS coronagraph performs best at short 
wavelengths.  The NICMOS coronagraph itself reduces the intensity of 
the background light by factors of a few to about ten within a few 
arcsec of an occulted star.  The unapodized edge of the 0\farcs3 
radius coronagraphic hole, however, acts both as a diffracting 
aperture and as a scattering surface caused by micro-roughness in 
manufacture, and is the primary source of instrumental contamination.  
The FWHM of the PSF (which scales linearly with wavelength) is 
$\sim$0\farcs12 at 1.1~$\mu$m and, hence, the F110W PSF is well 
contained within the hole.  When a star is centered in the 
coronagraph, F110W light from just beyond the second Airy minimum of 
the PSF falls on the hole-edge and is reduced in intensity by a factor 
of $\sim$250 with respect to the peak.  This is $\sim$6 times 
greater reduction in relative intensity at the hole-edge with respect 
to the peak than is obtained in H-band with the F160W filter.  
Thus, to maximize the disk-to-star contrast ratio by reducing the 
background scattered and diffracted light, the F110W filter is the 
spectral element of choice.  Our longer wavelength observations 
were planned assuming a disk surface brightness suggested by TB98.
It appears, however, that the scattered-light disk must be 
substantially fainter than had been reported (see \S 7.1).  
Hence, our longer wavelength images were of insufficient sensitivity to 
meaningfully contribute to an investigation of the disk reported 
by TB98 and TBR2000, and we do not discuss them further.

Because of the brightness of 55~Cnc itself relative to a surrounding 
dust disk, a reference PSF must be subtracted in order to reveal any 
low contrast circumstellar flux.  The efficacy of subtracting NICMOS 
coronagraphic PSFs from coronagraphic targets has been discussed at 
length (e.g., \citet{Schneider98a}).  The NICMOS PSF can exhibit small 
changes in the mid and high frequency structures on multi-orbit time 
scales which result from thermal instabilities in the heating of the 
HST secondary mirror support \citep {Bely} and from very small 
movements of the cold mask with respect to the detector 
\citep{Krist}.\footnote{The HST+NICMOS PSF is complex, in part due to 
a misalignment of the Lyot stop at the cold pupil, leading to 
striated structures in the energy redistributed radially along the OTA 
spider vanes and secondarily diffracted by the edges of the spider 
mask.  Because of small (sub-pixel) instabilities in the position of 
the cold mask with respect to the HST pupil (know as ``wiggles'', see 
\citet{Krist}) on both sub-orbit timescales and with secular drifts 
over weeks or months, the energy along the spikes is not always 
completely nulled by subtracting a reference PSF taken either at a 
different phase of the orbit or at a different epoch.  The residual 
energy along the spikes is radially distributed, essentially, as the 
difference of two J2 Bessel functions.  These often exhibit 
modulations in intensity over spatial scales of tens of pixels arising 
from very small secular variations in the OTA focal length, due to 
uncompensated desorption of the optical bench, and from sub-orbital 
changes due to ``breathing'' (see \citet{Bely}).  Thus, it is not 
always possible to null the integrated energy along the 45\degr\ and 
225\degr\ diffraction spikes simultaneously, nor to zero the total 
energy, particularly when a PSF taken at epoch different from the 
target is applied.  As this would bias any azimuthal averaging, we 
mask these regions in making all circumferential measurements 
discussed in this paper.} The effects of these variations may be 
minimized by obtaining a reference PSF within an interval shorter than 
the thermal time constant of the HST Optical Telescope Assembly (OTA).  
To this end we imaged HD~75216, a nearby bright star of similar color, 
to serve as a primary reference PSF. These data were obtained at 
nearly the same spacecraft attitude and in the same visibility period 
as the second of two contiguous orbits of 1.1-$\mu$m observations.

The HD~75216 images were reduced and calibrated in the same manner as 
the 55~Cnc images.  The photometrically calibrated PSF image was 
subtracted from the 55~Cnc images at each of the two field 
orientations after spatial registration and flux-scaling.  
Coronagraphic image registration was accomplished by means of cubic 
convolution interpolation \citep{Park} using the IDP3 program 
\citep{Lytle} developed by the NICMOS IDT. The target locations 
``behind'' the coronagraph were ascertained by measuring the 
unocculted target centroids (via Gaussian profile matching) in the 
calibrated acquisition images, and then applying the downlinked 
acquisition slew offsets executed by the NICMOS flight software in the 
coronagraphic acquisition process.  Flux-scaling of the PSF was 
accomplished by iteratively nulling the diffracted energy in the 
difference images along the OTA spider vanes while demanding that the 
adjacent background at large radii remain zero.  The sensitivity of 
the residual background to the PSF flux scaling is discussed by 
\citet{Weinberger99}, and we elaborate further here in \S 6.

The resulting pair of PSF-subtracted images of 55~Cnc is shown in 
Fig.~1.  The image in the left panel (a) was obtained at a spacecraft 
orientation angle of 42\fdg5 (1st orbit, Visit 60), such that the 
major axis of the disk reported by TB98 would lie essentially along 
the vertical.  In the right panel (b) the field was rotated 15\degr\ 
``clockwise'' around the occulted star.  Image artifacts in the 
optical system arising from scattered and diffracted starlight rotate 
with the detector and are essentially invariant.  While such artifacts 
remain spatially fixed, any circumstellar light rotates with the 
field, and hence an asymmetric disk would appear to rotate 15\degr\ 
about the target between image (a) and image (b).  The region interior 
to a distance of $\sim$25~AU (26.5 pixels, or $\sim$2\arcsec), 
represented by the green circles in Fig.~1, is largely dominated by 
residual scattered light from the star.  This is just interior to the 
inner boundary of the radial region fit by TB98, beyond which they 
report confidence that their data were unaffected by the presence of 
the 1\farcs5 radius occulting mask in the CoCo instrument.  The 
dashed yellow circle corresponds to a radius of 42AU, the outermost 
region for which TB98 confidently reported measuring a 1.6-$\mu$m flux 
excess.  Hence, the annulus between the green and yellow circles is 
the region considered in detail by TB98, in which they show an 
asymmetric disk with an r$^{-5}$ power law drop off in the surface 
brightness profile.

\section{Selection of Reference PSFs}

The measurement uncertainties in the circumstellar flux densities 
are dominated by image artifacts in the imperfect subtractions of 
reference PSFs. These uncertainties set the detection limits for a 
disk and are a strong function of radial distance from the 
occulted star. Near the star the NICMOS measurements are dominated 
by the residual systematic error in nulling out the imperfectly 
matched PSFs. While the 1-$\sigma$ pixel-to-pixel noise in spatial 
regions far from the influence of circumferential diffracted and 
instrumentally scattered light is typically $\sim$0.2~$\mu$Jy 
arcsec$^{-2}$, this is not germane in assessing the detectability 
of a disk where the sensitivities are ultimately limited by the 
much larger systematics in the presence of the PSF subtraction 
residuals. 

We have assessed our detection limits for our 55~Cnc images in two 
ways, as we have for other targets in the NICMOS IDT Environments of 
Nearby Stars (EONS) program \citep{Schneider98a}.  First we looked at 
the noise statistics of the circumstellar backgrounds following the 
registration, flux scaling and subtraction of a suite of well-matched 
reference PSFs.  We also examined the dispersion in the flux densities 
in equal-radii annular zones measured from each of the PSF-subtracted 
images.  Second we implanted a model disk with the characteristics 
described by TB98 into PSF subtracted target and flux-scaled reference 
images to see how well a disk of this nature would have been detected.  
To validate our results, in addition to applying the contemporaneous 
HD~75216 reference PSF that was obtained specifically for this 
purpose, we used 11 other PSFs obtained in the same manner in our 
observing program (see Table~3).  We restricted our reference PSFs to 
stars observed within ten weeks of the 55~Cnc observation to obviate 
any concerns of small secular instabilities in the NICMOS instrument 
over longer time scales.  Most of these stars were themselves 
candidates for possessing scattered light disks (with the exceptions 
of HD~75216 and HR 4748).  None of these, however, have shown any 
evidence of scattered light flux excesses in the spatial region of 
interest and hence were retained as reference PSFs.  In particular 
$\epsilon$~Eridani and $\tau^{1}$~Eridani were excellent choices as 
they were both significantly brighter than 55~Cnc (yielding very high 
S/N PSFs) and observed at more than one spacecraft orientation angle.  
We hasten to note that, while $\epsilon$~Eri has a known excess in 
the thermal IR and a dust ring detected in the submillimeter, this is 
at very large angular radii ($\sim$20-30\arcsec) as reported by 
\citet{Greaves}, and thus far outside the region of interest.

We were unable to detect any statistically significant circumstellar 
flux in either of the 55~Cnc images with any of the twelve reference 
PSFs used.  We consider separately the circumferential flux reported 
in the 28-42~AU annular zone by TB98 which they fit to an r$^{-5}$ 
power law, and the more complex surface brightness profile and peak 
flux density reported in the 19-28AU zone by TBR2000.  For both zones 
the contemporaneous HD~75216 PSF worked best in minimizing the 
subtraction residuals and introduced the least amount of structural 
artifacts in both orientations of 55~Cnc images (visits 60 and 61) at 
radii greater than $\sim$25AU, yielding nearly identical null results 
(see~\S6).  Similar null results, although with somewhat less 
statistical significance, were obtained using the other reference PSF 
stars listed in Table~3, where stars with two visit numbers on the 
same date were observed at two spacecraft roll angles.  We discuss in 
detail the results obtained using four observations of the first three 
stars (of spectral types F-K to minimize small color effects within 
the bandpass of the F110W filter) listed in italics in Table~3.  We 
compare these to a very similar null result obtained using HR 4748.  
This star was chosen because it has no evidence of, nor any 
expectations for, either a surrounding disk or detectable companions.

\section{Subtraction Sensitivities to Stellar Flux Estimates}

Potentially, circumstellar brightness profiles measured from 
PSF-subtracted images may be affected by either under or over 
subtracting the reference PSF. Aperture photometry of PSF cores of 
coronagraphically occulted target and reference stars is not possible.  
Thus, we measure the flux densities in many corresponding regions 
along the un-occulted diffraction spikes of the target and reference 
star PSFs at radii beyond where a disk may be present to establish the 
stellar flux density ratios.  For well exposed, contemporaneous, 
images the photometric flux density ratios derived from this approach 
are very robust.  Typically, we find that the dispersion in the 
measurements from different positions along the four diffraction 
spikes lead to uncertainties in the flux ratios of $<$1\%.  The 55~Cnc 
(Visit~61) and HD~75216 (Visit 62) images, shown in Fig.~2, were 
obtained closely together in the same HST orbit and at the same 
spacecraft roll orientation.  The 1-$\sigma$ uncertainty in the 
determined flux ratio of 2.51:1 is 0.6\% (6~milli-magnitudes).  This 
process does not work as well when using PSFs taken at different 
epochs, due to small changes in the PSF structures from secular 
instabilities in the HST focus and the NICMOS/HST optical interface.  
In such cases, and for the other PSFs employed as given in Table~3, 
errors of up to a few percent are typical.  As a check, we required 
that the flux ratios agreed with expectations based upon aperture 
photometry from unocculted target acquisition images transformed to 
the F110W band within their measurement and transformation 
uncertainties.

The common problem arising from misestimates in stellar flux ratios 
is the introduction of artifacts in PSF-subtracted images.  Such 
artifacts induce departures in the morphology and amplitude of the 
circumstellar light arising from the combination of diffraction and 
instrumental scattering due to a point source (i.e., a star without a 
disk).  If the brightness of the subtracted reference PSF is 
overestimated, an artifact resembling a circumstellar disk may arise.  
We quantify the magnitude of this effect for properly PSF-subtracted 
NICMOS F110W coronagraphic images of unresolved point sources in the 
2\farcs3 $<$~r~$<$~3\farcs3 annular zone where 50\% of the total 
area, along the orthogonal diffraction spikes, has been excluded with 
an oversized mask.  Typically, a 1\% change in the brightness of a 
flux-matched reference PSF of flux density, F, will result in a change 
in the integrated flux density of the instrumentally scattered light 
of $\sim$1.5x10$^{-5}$F into this region.

Here we consider the inverse possibility, that the reference PSFs 
employed and/or their flux density scaling, were biased in 
such a way as to suppress the visibility of intrinsic circumstellar 
scattered radiation.  Therefore, to assess the efficacy of the PSF 
subtractions for 55~Cnc, we performed identical image reductions and 
subtractions for control stars (i.e., stars without known or expected 
disks) whose F110W flux densities were renormalized to that of 55~Cnc 
($\sim$46~Jy), to serve as comparative nulls.  Reference PSFs were 
flux-scaled to the re-normalized null-stars by simultaneously 
demanding that the variance of the residual flux in the 
diffraction-spike masked images was minimized while the total flux 
density approached zero within the uncertainties imposed by estimating 
the flux ratios from the diffraction-spike fluxes.

We then estimated the sensitivity of the circumstellar-flux nulling 
process to misestimates in the stellar flux by adjusting the scaling 
factor of the reference PSFs by known amounts and then seeing how that 
affected the circumferential flux densities.  As an example, in Fig.~3 
we show three PSF subtractions of HR 4748, a control star with no 
detectable disk or companions, using HD~75216 as a reference PSF. 
Fig.~3a is the best result, having used the flux-matching criteria 
discussed.  The morphology of this image may be compared with the 
PSF-subtracted images of 55~Cnc in Fig.~1.  A small (and 
inconsequential) ``softening'' of the HR 4748 image is noticeable due 
to a minor focus shift in the HST/OTA between the two observational 
epochs, and this effectively smoothes the image over a scale of 
$\sim$2 pixels.  Nonetheless, the instrumentally scattered and 
diffracted light in the properly flux-scaled image is highly spatially 
correlated with the 55~Cnc images, suggesting the absence of any 
circumferential flux due to a disk.

As an example of flux mis-scaling, Figs.~3b and 3c show the effects of 
increasing, and decreasing, the flux density of the HD~75216 reference 
PSF by 10\% (an arbitrary illustrative amount) before subtraction.  
The resulting under and over subtractions in the PSF-subtracted images 
of HR 4748 are readily apparent, not only in the inner regions in 
these images ($<$~2\arcsec\ from the star), but also further out in 
the regions of particular interest (2\farcs3~$<$~r~$<$~3\farcs3) 
in the corresponding radial profiles.  The total flux density measured 
in this diffraction-spiked masked annular zone (corresponding to the 
28 to 42 AU region around 55~Cnc discussed by TB98 ) is 
0.52$\pm$0.36~mJy (standard deviation of all pixels) for the ``best'' 
subtraction for HR 4748.  This imperfect nulling implies there is a 
small bias due to unresolved systematics in this subtraction.  In the 
case of the 10\% under-subtraction, we find a flux excess artifact of 
5.61$\pm$0.40~mJy, and a deficit of -4.59$\pm$0.38~mJy for the 10\% 
over-subtraction.  These flux density artifacts imply a scattered 
light rejection ratio of $\sim$1x10$^{-4}$:1 into the masked annulus 
for the 10\% misestimates of the reference star flux densities.  This 
approximately agrees with the expectation for 1.5x10$^{-5}$ rejection 
into this annulus from a 1\% mis-scaling for typical coronagraphic 
performance.

In the case of the PSF-subtracted image of 55~Cnc (Visit 61) using 
HD~75216 (the same reference PSF as for the above null source), we 
measured an integrated flux density in the masked 28 to 42 AU region 
of +0.40~mJy after correcting for the masked 
area (\S~7.1).  Changing the flux density of HD~75216 
in the subtraction by $\pm$0.6\%, the 1-$\sigma$ uncertainty in the 
flux ratios, gives rise to uncorrected flux density measurements of 
+0.21 and +0.18~mJy.  But, the TB98 results would have predicted a 
flux density in the masked annulus of 4.0~mJy, a factor of $\sim$20 
higher than seen in our Visit 61 image.  Suppression of 3.8~mJy of 
flux in the masked annulus from a 46~Jy source due to a flux 
mis-scaling of the reference PSF would have required a photometric 
error of $\sim$8\% in establishing the 55~Cnc:HD~75216 F110W 
brightness ratios, a factor of 13 higher than our estimated 
uncertainty.  For the 55~Cnc (Visit 60) image, taken in an earlier 
orbit at a different spacecraft roll orientation, the flux ratio to 
the same reference PSF was found to be 2.52$\pm$.023 (0.9\% 
1-$\sigma$).  The increase in the uncertainty in the flux-scaling by 
50\% was due to secular changes in the 55~Cnc PSF between the two 
orbits.  These data resulted in a higher integrated flux density of 
+0.32 mJy measured in the masked annulus, but still an order of 
magnitude lower than suggested by TB98.  These measures are consistent 
with the typical scattered light rejection properties of the NICMOS 
coronagraph at 1.1$\mu$m and strongly suggest that we did not 
inadvertently null a bright disk by oversubtracting reference PSFs.

Finally, we examine the radial profiles resulting from under, over, 
and properly scaled subtractions of the reference PSF (HD~75216) from 
the null source (HR 4748) and compare them to the r$^{- 5}$ profile 
suggested for 55~Cnc by TB98 (green line in Fig.~3).  We note that it 
is possible to null the integrated flux density of a TB98-like disk by 
severely oversubtracting a reference PSF. However, the radial 
morphology of the disk-like artifact produced by an undersubtraction 
of the same magnitude has a complex radial morphology which bears {\em 
no resemblance} to an r$^{-5}$ dependency in the same angular radial 
zone considered by TB98.

\section{Flux Densities}

High spatial frequency radial artifacts (``streaks'') in the PSF- 
subtracted images, at azimuthal scales comparable to single pixels, 
contaminate the images at small radii.  As the error in the background 
measurement at a given radial distance is due to azimuthal variations 
in intensity, these well known image artifacts cause an 
over-estimation in the formal measurement error.  To mitigate the 
radial artifacts to some degree, we measure the brightness profiles 
(discussed in \S~7.1 and 7.2) after applying a 3x3 pixel smoothing 
kernel to the difference images.  This spatially degrades the 
resolution by a factor of 2 to $\sim$0\farcs23, but smooths over the 
high spatial frequency artifacts, thus yielding a better estimate of 
the uncertainty in the larger scale backgrounds at smaller radii.

\subsection{The 28--42 AU Zone} \label{zone1}

For each of the PSF subtractions we measured the total flux density 
enclosed in an annulus between 28 and 42~AU, the annular zone fit by 
TB98.  In doing so we excluded pixels along the diffraction spikes and 
in two small regions where scattered light artifacts are known to be 
induced by the coronagraphic optics.  The resultant region where 
observations were made included 1495 pixels covering 8.6~arcsec$^{2}$.  
We correct the measured total flux densities in the masked annulus by 
a factor of 2.06, linearly scaling the measured flux by the ratio of 
areas in unmasked to masked annuli.  The integrated flux densities, 
corrected for the unsampled areas, are presented in Table~4 along with 
their error estimates (standard deviation over all pixels) for each of 
the subtractions.

Our ability to measure the integrated flux density in this region has 
an internal accuracy of $\sim$0.08~mJy when subtracting any of the 
non-contemporaneous PSFs, or $\sim$0.05~mJy when using the HD~75216 
PSF obtained contiguously with the 55~Cnc observations.  However, as 
is evident, the flux densities measured amongst these representative 
PSF subtractions have a dispersion of about $\pm$0.9~mJy.  This is 
indicative of the fact that, as an ensemble, the measures are 
dominated by systematics in the subtractions themselves.  These 
systematics arise, primarily, from uncertainties in the flux-scaling 
of the reference PSFs, but also from structural variations among the 
PSFs used.  The HD~75216 (Visit 62) PSF was taken in the same target 
visibility period (shorter than the thermal time constant of the 
NICMOS + HST focal plane variations) as the Visit 61 images of 55~Cnc.  
The resulting difference image is the most free of optical artifacts 
at distances $\gtrsim$~2\arcsec\ induced by PSF mismatches (see \S~6).  
The uncertainty in the flux-scaling of this PSF to the 55~Cnc images 
(both visits) was smaller by a factor of two than for any other PSF. 
Thus, a priori, we expected this to be the ``best'' matched PSF (least 
devoid of systematics), and this was found to be the case based upon 
the criteria already discussed.  Given the systematic uncertainties, 
we cannot say if the +0.4 mJy integrated density of the circumstellar 
flux seen in the Visit 61 - Visit 62 image is wholly instrumentally 
scattered flux, or if there is an underlying disk component which may 
be below our detection threshold.  Allowing for the 0.6\% uncertainty 
in the flux-scaling and the internal measurement error of 0.06~mJy 
(see Table~4), we would be unable to detect a disk-integrated flux 
density of $\lesssim$~0.4$\pm$0.07~mJy.  The F110W flux density of 
55~Cnc itself is $\sim$46~Jy, implying a instrumental scattered light 
rejection ratio of $\sim$10$^{5}$ within this radial zone for these 
observations.

In Fig.~4 we show the azimuthally averaged surface brightness 
profiles for the two 55~Cnc images using HD~75216, $\epsilon$~Eri and 
$\tau^{1}$~Eri (at two spacecraft roll angles observed contiguously in 
one visibility period) as reference PSFs.  These profiles exclude the 
imperfectly nulled diffraction spikes, which are masked in the 
accompanying images.  The surface brightness profiles, measured in 1 
AU wide radial increments, were determined after modestly smoothing 
the PSF-subtracted images as discussed in \S~7.  The 1-$\sigma$ error 
bars indicate the standard deviation of all unmasked pixels in each 
incremental radius about the mean value at that radius, $\sigma$(N), 
and not the standard deviation of the mean, $\sigma$(mean).  If the 
data in each pixel in an annulus were uncorrelated (i.e., independent) 
then $\sigma$(mean) would simply be $\sigma$(N)/$\sqrt{N}$.  Image 
artifacts at spatial frequencies of several pixels result in partial 
correlations of measured intensities in adjacent pixels in the 
difference images, particularly at smaller radial distances to the 
star.  Thus, $\sigma$(mean) approaches $\sigma$(N)/$\sqrt{N}$ at large 
radii, but becomes more uncertain at smaller radii in a complex 
manner.

We have characterized the major axis flux densities for the disk 
suggested by TB98 (green lines in Fig.~4) as F(r) = 1.24 
(r/28~AU)$^{-5}$~mJy arcsec$^{-2}$ where we have renormalized the TB98 
H-band spectral flux densities to compare with our NICMOS F110W 
measures.  Here we assume neutral gray scattering by the circumstellar 
grains at these wavelengths and hence the dust would take on the color 
of 55~Cnc.  We believe this is a reasonable assumption for dust debris 
particles, which are likely to have a characteristic size of at least 
a few microns.  We have found neutral scattering in other debris disks 
observed by NICMOS in the F110W and F160W (very close to H) 
photometric bands (e.g., HR~4796A, \cite{Schneider99}; HD~141569, 
\cite{Weinberger99}, and \citet{Augereau}; and TW~Hya.  
\cite{Weinberger00}).  We estimate the NICMOS F110W:H spectral energy 
density ratio of 55~Cnc using the STSDAS CALCPHOT routine to be 
$\sim$1.20 $\pm$ a few percent given the inherent uncertainties in 
the color transformation.  In doing so we employ a Kurucz model 
atmosphere with an effective photospheric temperature of 5336K, 
log(g)~=~4.3, and a metalicity of [0.45]~dex above solar all in 
accordance with both \citet{Gonzalez98a} and \citet{Fuhrmann}.

Integrating the reported r$^{-5}$ profile (TB98), the expected total 
flux density in this region from a TB98-like disk is 8.2~mJy.  This is 
a factor of 20 times higher than the $\sim$0.4-mJy detection limit 
that we derived using our best reference PSF (HD~75216), or more 
conservatively an order of magnitude higher than a limit based on the 
variations from the ensemble of PSF-subtracted images.  We rule out 
the possibility of not detecting this much flux due to 
over-subtraction of our PSF stars, as detailed in \S~6.

\subsection{The 19--28 AU Zone} \label{zone2}

TBR2000 report measurable flux densities to within $\sim$19~AU of 
55~Cnc, at the inner usable radius of CoCo system, which were not 
presented by TB98 in their initial analysis of the same data.  TBR2000 
more recently inferred a peak H-band surface brightness of the 
circumstellar light of $\sim$3~mJy arcsec$^{-2}$ near the edge of the 
1\farcs5 radius coronagraphic mask at $\sim$20'~AU (Fig 3 of 
TBR2000).  Although the TB98 power-law fit indicates that the surface 
brightness is apparently continuing to rise as r$^{-5}$ toward smaller 
radii there is an obvious roll-off in the surface brightness profile 
at r~$<$~28AU, as presented by TBR2000.  We reproduce their measured 
radial surface brightness profile (renormalized to the NICMOS F110W 
band as discussed in \S~7.1) along with an extrapolation of the 
suggested r$^{-5}$ profile for radii between 19 and 30~AU in Fig.~5.  
An abrupt discontinuity in the TBR2000 surface brightness at 
r~$\approx$~21~AU is also seen where the circumferential flux density 
is an order of magnitude lower than the $\sim$3~mJy arcsec$^{-2}$ 
reported at 20~AU (or $\sim$4~mJy~arcsec$^{-2}$ in NICMOS F110W).  It 
is difficult to attribute this large amplitude variation as being 
intrinsic to the source, as it occurs over a much smaller spatial 
scale than the size of the best IRTF/CoCo seeing disk of 0\farcs4 
(5~AU at the distance of 55~Cnc) reported by TBR2000.  We suggest that 
this discontinuity, and the reported $\sim$3~mJy arcsec$^{-2}$ H-band 
peak, might arise from scattered light artifacts within the CoCo/IRTF 
optical system.  Neither of these are seen in the NICMOS data where 
the size of a resolution element in the camera 2 F110W coronagraphic 
PSF is 0\farcs12 ($\sim$1.6 AU) sampled with 0\farcs076-square 
pixels.  TRB2000 co-present a similarly-observed radial brightness 
profile for $\upsilon$~Andromedae as a comparative null.  We note, 
however, a very similar rise toward smaller radii in the 
$\upsilon$~And circumferential brightness profile at nearly the same 
angular radial distance and magnitude as indicated for 55~Cnc.  
Unfortunately, TBR2000 do not present the data in the null profile at 
radii $<$ 28AU, which makes comparing the morphology of the brightness 
profile of their control star to that of 55~Cnc difficult in the 
regions of highest reported circumstellar flux.

In the 25-30~AU region, where the TBR2000 measurements begin to 
diverge from the TB98 r$^{-5}$ power law fit, we obtain a null result 
in detecting a disk to the levels of our sensitivity imposed by the 
systematics with flux densities of 0.17$\pm$0.22 mJy arcsec$^{-2}$ at 
25~AU and 0.04$\pm$0.12~mJy arcsec$^{-2}$ at 30~AU. At the same 
distances TBR2000 suggest brightnesses of $\sim$2 and 
1~mJy~arcsec$^{-2}$, respectively.  These flux densities exceed the 
NICMOS determined upper limits by factors of 12 and 25 in the same 
radial zones, respectively, as we also saw for the integrated flux 
densities in the 28~$<$~r~$<$~40~AU region.  This suggests that, if a 
disk with flux densities reported by TB98 and TRB2000 was present in 
the 19-28~AU annular zone, it should also have been readily visible in 
the NICMOS images.

\section{Model Disk Implantation}

Since we failed to find any circumstellar light above the level of the 
instrumental artifacts in the (imperfect) PSF subtractions, we then 
investigated how readily a disk with the photometric properties and 
morphology suggested by TB98 could have been detected in our images.  
To assess this we built a simple model disk characterized as described 
in \S~7.1 and consistent with the TB98 results.  To replicate the 
suggested major-to-minor axis flux density ratio of 1.7:1 we 
sinusoidally varied the intensity of the disk azimuthally.  We then 
implanted a registered noiseless image of the model disk, with the 
major axis at a PA = 42\fdg5 (vertical in our Visit 60 image and 
corresponding closely to the TBR2000 indicated orientation), into our 
PSF subtracted images of 55~Cnc.  After implantation, we fully 
recovered the 8.2 mJy of superimposed flux from the model disk 
measured in the same manner as the PSF-subtracted images of observed 
stars.  This is a factor of about 20 higher than the measures using 
our contemporaneous PSF subtractions.  The model disk implantation 
suggests that a comparable real disk in the 55~Cnc images should have 
been seen with a S/N of about 65.  This recovered flux density is also 
about 10 times higher than the dispersion in the scattered-light 
background amongst all our measures (Table~4).  Given this, we would 
have expected to detect an integrated flux excess from the suggested 
disk in our PSF subtracted images.

The azimuthally averaged brightness profiles were also measured from 
the disk-implanted images, and their statistical errors were assessed.  
In all cases the model disk was recovered in the presence of the noise 
in the observations at the levels indicated by the superposition of 
the measured subtractions as given in Table~4 with a disk of 
integrated flux density 8.2~mJy.  For example, in Fig.~6 we show this 
for the subtraction of the Visit~60 image of 55~Cnc using the HD~75216 
reference PSF for comparison with the null results already discussed 
in concert with Fig.~4.  The implanted disk with an integrated S/N 
of $\sim$80 is readily apparent.  As for Fig.~4, the error bars for 
Fig.~6 indicate the standard deviations of all pixels about the mean, 
$\sigma$(N), of each 1~AU wide annulus ($\sim$2 to 1~$\sigma$(N) for 
28 to 42~AU, respectively).  If the residuals in the subtracted data 
smoothed with a three pixel kernel are otherwise spatially 
uncorrelated, the standard deviations of the means, $\sigma$(mean), 
would be reduced to $\sim$$\sigma$(N)/9 in the 28 to 42~AU annular 
zone, giving rise to 18 to 9~$\sigma$(mean) detections in each of the 
corresponding zones.  This is consistent with the assessment of our 
detection limits based upon the systematics in the PSF subtractions 
detailed in \S~7.1.

\section{Azimuthal Anisotropy}

TB98 reported an anisotropy in the surface brightness of the putative 
disk manifesting itself as a 1.7:1 ratio in flux densities measured in 
sectors oriented along the disk major axis (aligned at a position 
angle of $\sim$45\degr\ ) compared to the minor axis flux.  In our 
Visit~60 image of 55~Cnc the major axis of the reported disk lies 
essentially in the vertical direction.  Thus, if a disk as bright as 
suggested by TB98 was present, we could have observed a significant 
difference in the major to minor axis integrated flux densities.  
However, as the scattered-light component of the 55~Cnc disk actually 
appears to be at or below our detection limits, our data cannot 
meaningfully contribute to investigating the axial asymmetry suggested 
by TB98.  We note, however that the azimuthal morphologies of our two 
55~Cnc observations (Visits 60 and 61), which were rotated 15 degrees 
with each other, are extremely similar (see Fig.~1) and both very 
closely resemble our null-star subtractions (for example, inset, a, of 
Fig.~3).  The lack of differential azimuthal asymmetry in the two 
differently oriented 55~Cnc images, and their similarities to the 
comparative null images, further suggest that the integrated flux 
density we measure at the $\sim$0.4~mJy level (for our best PSF 
subtraction in the 28-42~AU zone) is likely instrumental in origin.

\section{Comparison with Other NICMOS Observations}

While the r$^{-5}$ power law index reported for the putative 55~Cnc 
scattered light disk by TB98 is rather steep, the flux densities in 
specific regions at different radii are comparable to (and in some 
cases larger than) the positive NICMOS detections of material around 
other IR excess stars.  In Fig.~7 we compare the PSF-subtracted 
NICMOS coronagraphic image of the circumstellar environment of 55~Cnc 
(the null detection from Visit 60 using HD~75216 as a PSF) to two 
other sources of similar spatial scales and surface brightnesses; 
TW~Hya \citep{Weinberger00} and HD~141569 \citep{Weinberger99}.  
The data from which these images were obtained was reduced and 
processed in a manner similar to the 55~Cnc data.  And, like 55~Cnc, 
several PSF stars were applied to each, yielding consistent 
morphological and photometric results.  We present these images, along 
with quantitative assessments of our detection limits, as a 
demonstration of the efficacy of the NICMOS coronagraphic system in 
detecting and spatially resolving reflection disks with properties 
similar to those suggested for 55~Cnc by TB98.

From the TB98 results we would have expected an F110W azimuthally 
averaged surface brightness of 1 mJy arcsec$^{-2}$ at a radius of 
$\sim$28AU, (29 pixels in NICMOS camera 2).  In panel, a, of Fig.~7 
(55~Cnc image) we indicate where this region is located.  The dynamic 
range of this display is the same as that of the 55~Cnc image and the 
comparative images in panels, b and c, with the high end (white) set 
to the anticipated level of the TB98 disk at 28AU 
(1~mJy~arcsec$^{-2}$).  As may be inferred from the analysis of our 
detection limits and model disk implantation experiment, this is 
rather far out from the residual glare of 55~Cnc, and we would expect 
a 1~mJy~arcsec$^{-2}$ disk to have been readily seen at this radius.  
The zonal regions of comparable surface brightness in the TW~Hya image 
and the inclined ring of maximum brightness around HD~141569 are 
closer in, in angular distance, primarily because of the greater 
distances to these stars.  Since the apparent brightness of 55~Cnc 
exceeds both TW~Hya and HD~141569 by at least an order of magnitude 
the residual amplitudes in the image artifacts around 55~Cnc at 
comparable radii are correspondingly larger.  The amplitude and 
morphology of the scattered light residuals are very similar to those 
seen in the mutual subtractions of unresolved point sources of 
comparable brightness.

\section{Discussion}

Our non-detection of near-IR scattered light from the 
circumstellar debris system in 55~Cnc is inconsistent with the 
flux densities and morphology of the surface brightness profile 
reported by TB98 and TBR2000. The discrepancy is significant at 
radii $\gtrsim$~28~AU as discussed by TB98. It is also significant 
in the 19-28~AU zone more recently presented by TBR2000. Given the 
NICMOS 1.1-$\mu$m sensitivities and detection limits, this interior 
region is effectively probed to a limiting distance of $\sim$15~AU 
with the NICMOS coronagraph, yet we detect no statistically 
significant circumferential flux excess in any of these regions.

\subsection{Disparity in Scattered Light} \label{light}

NEAR-IR. The NICMOS upper detection limits to the integrated scattered 
light at 1.1$\mu$m are 10 to 20 times lower than the 
8.2~mJy~arcsec$^{- 2}$ level in the 28-42 AU circumstellar region 
suggested by TB98.  Additionally, in the region interior to 28AU as 
discussed by TBR2000, despite the presence of image artifacts induced 
by the HST+NICMOS optics close to the star, the NICMOS system 
sensitivity was sufficiently high to have detected a disk with a 
surface brightness at least ten times fainter than reported.  Yet, 
none was seen.

MID/FAR-IR. The mid/far-infrared optical depth for the disk around 
55~Cnc is measured at $\sim$6x10$^{-5}$ (\citet{Dominik}; 
\citet{Jayawardhana}).  If the dust in the 55~Cnc disk is like that 
found around $\beta$~Pictoris, HR~4796A and HD~141569, the albedo of 
the grains would be in the range $0.2 \lesssim \omega \lesssim 0.5$.
Thus, the scattered-light optical 
depth at a given wavelength would be a few times smaller than the IR 
optical depth (e.g., for HD~141569 $L_{\rm IR}/L_{*}$ = 8x10$^{-3}$ 
and $F_{\rm scattered}/F_{*}$ = 3x10$^{-3}$ at 1.1~$\mu$m ).  For 
55~Cnc we would therefore predict $F_{\rm scattered}/F_{*}$ = 
1--5x10$^{-5}$ at 1.6~$\mu$m.  At H-band, 55~Cnc is 4.17~mag 
\citep{Persson} or 22.3~Jy, so the predicted disk flux would be 
$\sim$0.38~mJy.  The TBR2000 surface brightness profile indicates a 
minimum scattered-light flux density of $\sim$6.6~mJy (integrating 
outward from their limiting radius of 19AU), $\sim$17 times greater 
than expected, and implying $\tau_{\rm scatter}$ = 3x10$^{-4}$.  This 
scattering fraction is more than an order of magnitude larger than 
observed in other disks.  Assuming this large $\tau_{\rm scatter}$, the 
albedo inferred is 0.83.  This is not out of the realm of possibility 
for ice.  We note, however, that it is higher than the albedo for the 
disks around a) HR~4796A (0.2 combining \citet{Schneider99} and 
\citet{jura}), b) HD~141569 (0.3-0.5 from \cite{Weinberger99}), c) 
$\beta$~Pictoris (0.6 from \citet{Pantin}), and more than an order of 
magnitude higher then the very dark 0.06 assumed by TB98 for Kuiper 
belt-like objects.  While it is physically possible that such dust 
particles might exist, given the NICMOS results, it seems more 
reasonable that the flux density due to scattering by circumstellar
grains is actually an order of magnitude or more lower than 
reported by TB98.

FAR-IR/SUB-MILLIMETER.  The absence of scattered NIR radiation in the 
NICMOS images from circumstellar dust, to the flux density limits 
previously discussed, in the 19~$<$~r~$<$~42 AU the annular zone 
(reported on by TB98 and TBR2000) is consistent with a model of the 
disk developed by \cite{Dominik} based upon ISO flux excesses.
They found the 55 Cnc dust grains cannot exceed a temperature of 100K,
must be located at radii $\gtrsim$ 35 AU, and are likely distributed in a zone 50--60~AU from the star.  Fitting a blackbody spectral energy 
distribution to the far-IR and their more recent JCMT/SCUBA 450$\mu$m 
850$\mu$m observations, JAY2000 suggest a central hole with
r~$\lesssim$~35~AU for a dust temperature of 60K (consistent with \cite{Dominik}). They note that for grains as warm as 
100K the radius of the central hole could be as small as 13AU.
However, if the surface brightness reported by TB98 continued increasing
inward of 28~AU with their suggested r$^{-5}$ power law, the
expected 1.1-$\mu$m surface brightness at 19~AU 
(reported on by TBR2000) would be $\sim$8~mJy~arcsec$^{-2}$,
easily detectable, but not seen, with NICMOS (Fig.~5).  
Such an inward extrapolation is almost certainly not warranted as the 
scattering properties of the dust would have to be radially isotropic.  
This is very unlikely, as the albedo (for example) would have a strong 
dependence on radius as the ice sublimation temperature is reached 
\citep{Pantin}.  Nonetheless, in the presence of a sub-mm disk 
with flux densities discussed by JAY2000, some amount of excess 
scattered-light flux would be expected closer to the star, but this 
must be below our detection limits as discussed in \S~7.2.

\subsection{Resolving the Disparities} \label{resolving}

As expected from the observed thermal and submillimeter fluxes, we are 
unable to confirm the detection of scattered near-IR light reported by 
TB98.  There are two possibilities for our non-detection: 

1) We feel 
that the most likely possibility is that the TB98 and TBR2000 
detection is spurious.  In this case, our non-detection places an 
upper limit to the 1.1-$\mu$m emission of at most $\sim$10\% of the 
integrated spectral flux density in a disk with the properties 
reported by TB98.  This possibility relieves the one to two order of 
magnitude discrepancy in the estimates of the dust mass inferred by 
TB98 and JAY2000.  It also removes the need to invoke unusual grain 
properties and geometries in an attempt to explain the TB98 reported 
flux densities.  This conjecture is supported the fact that all three disks 
reported by TBR2000 have very similar morphologies and flux density 
distributions (i.e., surface brightness vs. radial distance in arc 
seconds) despite the lack of any thermal IR radiation in those systems 
as sampled by ISO.

2) The NICMOS data are faulty.  However, given the extensive 
discussions in \S~4-7, and our successful detection of disks of other 
systems of comparable brightness and spatial extent (e.g., Fig.~7), we 
reject this alternative.

\section{Conclusions}

Utilizing the NICMOS coronagraph on board HST, we have attempted to 
confirm and characterize the circumstellar debris disk surrounding 
55~Cnc seen in scattered NIR light as reported by TB98.  We find, 
however, no significant evidence for the existence of a near-IR 
excess, nor any morphological structures resembling a disk, in our 
highest sensitivity (F110W) PSF subtracted images.  The NICMOS 
observations indicate that any light scattered from a disk at 
1.1~$\mu$m must be at least ten (and likely twenty) times fainter than 
suggested by TBR2000.  The actual scattered flux could be even 
smaller, adding observational support to the conjecture by JAY2000 
that TB98 may have overestimated the surface brightness of a scattered 
light disk.  This conjecture should be tested by additional NIR 
observations designed to set more stringent limits on the brightness 
and morphology of a disk which may be seen in reflected light.

We are unable to account for the disparity in the NICMOS and CoCo 
observations.  We cautiously suggest that the flux excess seen by TB98 
may have been spurious as a result of the difficulties involved in 
ascribing the correct amount of flux to subtract via a scaled PSF. 
This interpretation is consistent with the conjecture by JAY2000 that 
TB98 may have overestimated the NIR disk brightness based upon their 
estimates of the total dust mass.  Further CoCo observations are 
required to resolve the disparity in these two data sets.  To help 
resolve this, we advocate cross-correlating NICMOS and CoCo 
coronagraphic data sets.  We suggest the need for observations of well 
characterized and independently confirmed NICMOS scattered light disks 
such as TW~Hya (also seen in the optical with WFPC-2 by \citet{Krist}) 
and HD~141569 (independently observed with NICMOS by \citet{Augereau}) 
with the CoCo system.

\acknowledgments

We gratefully acknowledge the many contributions to the other
members of the NICMOS Environments of Nearby Stars and Instrument
Definition Teams during the course of this investigation.
We also thank our program coordinator, Douglas Van Orsow, and contact 
scientist, Alfred Schultz, at STScI for the assistance in the 
last-minute scheduling and implementation of these observations.  This 
work is supported by NASA grant NAG~5-3042.  This paper is based on 
observations with the NASA/ESA Hubble Space Telescope, obtained at the 
Space Telescope Science Institute, which is operated by the 
Association of Universities for Research in Astronomy, Inc.  under 
NASA contract NAS5-26555.

\clearpage

\figcaption{PSF subtracted F110W coronagraphic images of the 
circumstellar region in the vicinity of 55~Cnc in two successive 
HST orbits with a flux re-normalized PSF of HD~75216 subtracted. 
The fields shown are 9" wide. a) North, in the left image, is to 
the upper right (42.5\degr\ from vertical, nearly along the 
diffraction spike mask). b) With the reorientation of the 
telescope, the field as has been rotated 15\degr\ clockwise with 
respect to the image in panel a. \label{fig1}}

\figcaption{NICMOS F110W coronagraphic images of 55~Cnc (a) and 
HD~75216 (b) flux-scaled by a factor of 2.51 by ratioing the 
intensities of the light from the diffraction spikes of (a) and (b) as 
described in the text.  Systematic instabilities give rise to slightly 
different structures in the two PSFs, resulting in the residuals seen 
in the difference image (c).  Panels a and b are stretched over a 
dynamic range of 0 to 0.1 mJy per pixel.  Panel c is a bipolar stretch 
from -0.1 to 0.1 mJy per pixel (same positive domain) to illustrate 
the nature of the subtraction residuals.  The images have been rotated 
45.5\degr\ counter-clockwise to place the diffraction spikes along the 
horizontal and vertical, so North is up and to the left in panel a.
\label{fig2}}

\clearpage

\figcaption{Incorrect estimates of target and reference star 
brightness affect the morphological structure and measured 
circumstellar flux in PSF-subtracted images. Inset a represents a 
``best'' image of the null source HR 4748, where the reference PSF 
(HD~75216, obtained contemporaneously with the 55~Cnc images) was 
subtracted by variance-minimized nulling of the circumferential 
flux (based upon measured stellar flux ratios checked with band-
transformed photometry from target acquisition images). Under and 
over subtracted images, by 10\% in the F110W reference star flux, 
are shown for PSF-subtracted images of HR 4748 (insets b and c). 
These PSF-subtracted images of the null source may be compared 
with those of 55~Cnc in Fig.~1. Azimuthally averaged radial 
profiles measured from these images are also shown and may be 
compared with the topmost panels in Fig.~4. Radial zones at the 
same angular distances considered for 55~Cnc by TB98 (and 
discussed in \S 7.1) are indicated by the yellow background. The 
green line indicates an r$^{-5}$ surface brightness profile 
suggested by TB98 and fit to the 55~Cnc data presented in
Fig.~3 of TBR2000 (renormalized to F110W as discussed in the text).
\label{fig3}} 
\clearpage

\figcaption{Azimuthally averaged F110W ($\mu$m) surface brightness
profiles\protect\footnote{In measuring these 
profiles we also masked small 
regions which are known to be repeatably affected by image 
artifacts in these radial zones at approximately ``11 o'clock'' 
and ``2 o'clock'' in the frame of the detector. The artifacts are 
obvious in the inset images (where they are unmasked only for 
illustrative purposes). In the $\tau^{1}$ Eri PSF subtractions 
they appear as negative, whereas in the $\epsilon$ Eri PSF 
subtractions they are positive, while they are not so obvious in 
the HD~75216 subtraction. The cause for the parity reversal is 
understood to be due to the ``breathing'' of the HST focal plane. 
The $\tau^{1}$ Eri and $\epsilon$ Eri PSFs were both slightly 
afocal with respect to the 55~Cnc image, but shifted in opposite 
directions, causing a phase inversion upon subtraction in these 
scattered light artifacts. Other implications for the effects of 
OTA focus changes due to breathing on NICMOS images are discussed 
in detail by \citet{Kulkarni}.}
measured from PSF-subtracted images of 55~Cnc at two field 
orientations using four of our twelve reference PSFs. The images 
shown in the insets were smoothed as described in \S 7.1 before 
computing the brightness profiles. The error bars indicate the 
standard deviations about the mean of all good pixels in each 1 AU 
wide annular zone The putative 55~Cnc disk is oriented vertically 
between the masked diffraction spikes in the Visit 60 image (left) 
and is rotated 15\degr\ clockwise in the Visit 61 image (right). 
The yellow regions correspond to the radial zones considered by 
TB98. As in Fig.~3, the TB98/TBR2000 suggested r$^{-5}$ power 
law fit, adjusted for the NICMOS F110W band pass, is indicated by the 
green line in each panel. The inset images from which the profiles 
were derived are stretched, scaled and annotated as in Fig.~1 
($\pm$ 2 mJy arcsec$^{-2}$, and 9" wide, 25AU and 42 AU zonal regions 
indicated). 
\label{fig4}} 
\clearpage

\figcaption{The 19-30 AU diffraction-spike masked azimuthally 
averaged surface brightness measurements from several NICMOS PSF 
subtracted images of 55~Cnc are presented along with the TRB2000 
profile (renormalized to F110W, see text \S 7.1) and an 
extrapolation of the TB98 r$^{-5}$ power law fit to their data at 
radii interior to 28 AU. For illustrative purposes we show the 
results for the Visit 60 image of 55~Cnc using the same reference 
PSFs as in Fig.~4, though similar results are obtained using our 
other previously discussed PSFs (Table~1) and for the Visit 61 
data. When applied to the Visit 60 data the Visit 62 PSF is a poor 
match at radii $<$$\sim$2\arcsec\ (dotted region, $<$ 23AU) due to the 
thermal history of the telescope assembly and the relative phasing 
of these observations with the HST orbit as discussed by 
\citet{Schultz} and give rise to the anomalous (and non-physical) 
negative residuals in this PSF subtraction at these radii which 
are rejected. 
\label{fig5}}

\figcaption{a) Scaled PSF subtraction of 55~Cnc (Visit 60) by HD~75216 
(Visit 62) in the annulus observed and fit by TB98 (28-42 AU zone).  
b) Image (a) smoothed to remove high frequency radially directed 
artifacts.  c) Axisymmetric r$^{-5}$ model disk with flux density 
characteristics described by TB98.  d) Model disk (c) registered and 
implanted in subtraction (a) with major axis oriented as discussed by 
TB98 (along vertical axis in these data).  e) Image (d) smoothed as 
for image (b).  f) Surface brightness profile of implanted model disk 
the region discussed and fit by TB98 in gray.  Error bars in (f) are 
the standard deviations of all unmasked pixels about the mean in 1 AU 
wide annular zones at the indicated radial distance.  Panels a-e are 
shown at the same linear stretch as indicated by the scale bar.  Data 
affected by HST diffraction spikes (along diagonals) were masked and 
excluded in measuring the surface brightness profile.
\label{fig6}}
\clearpage

\figcaption{Comparative imaging of circumstellar material at 1.1$\mu$m 
via PSF-subtracted NICMOS coronography.  a) 55~Cnc (H=4.17) 
non-detection.  b) Face-on disk about the classical T~Tauri star 
TW~Hya (H=7.34).  c) Inclined debris disk circumscribing Herbig AeBe 
star HD~141569 (H=6.89).  All panels are linear displays at the same 
stretch from -0.1 to 1.0 mJy arcsec$^{-2}$, and presented at the same 
angular image scale.  The red circles in (a) at a radius of 
2\farcs2 (28AU) and (b) at 1\farcs3 (65AU) respectively indicate 
the locations of the 1.0~mJy~arcsec$^{-2}$ surface brightness contours 
expected from TB98 in 55~Cnc, and observed for TW~Hya.  The bright 
ring at 1\farcs85 (185AU) in (c) has an azimuthally averaged 
surface brightness of 0.27~mJy~arcsec$^{-2}$.  The solid circles 
indicate the physical size of the NICMOS coronagraphic hole 
(0\farcs3 radius).
\label{fig7}}

\clearpage
\begin{deluxetable}{ccccc}
\tablenum{1}
\tablewidth{0pc}
\tablecaption{Spectral Filters and Photometric
Calibrations\label{tab_filters}}
\tablehead{
\colhead{Filter}
          &\colhead{$\lambda_{cent}$ ($\mu$m)}
                   &\colhead{FWHM ($\mu$m)}
                            &\colhead{mJy/ADU/s}
                                      &\colhead{0-
pt\tablenotemark{a}}
}
\startdata
F110W   &1.0985   &0.5920    &2.031    &1775 \\
F160W   &1.5960   &0.4030    &2.190    &1083 \\
F187N\tablenotemark{b}  &1.8740   &0.0192    &41.07    & 825 
\\
F204M   &2.0313   &0.1050    &7.711    & 735 \\
F237M   &2.3978   &0.1975    &4.913    & 635 \\
\enddata
\tablenotetext{a}{Janskys for 0 Vega magnitude}
\tablenotetext{b}{Used for Target Acquisition}
\end{deluxetable}

\clearpage
\begin{deluxetable}{ccccccccc}
\tablenum{2}
\tablewidth{0pc}
\tablecaption{NICMOS Coronagraphic Exposure Log\label{tab_log}}
\tablehead{
\colhead{Target}
        &\colhead{Filter}
                &\colhead{UT Start}
                          &\colhead{Sequence}
                                   &\colhead{Reads}
&\colhead{N\tablenotemark{a}}
    &\colhead{Exp (s)}
            &\colhead{Integ (s)\tablenotemark{b}}
                      &\colhead{Orient 
($^\circ$)\tablenotemark{c}}
}
\startdata
55~Cnc  &F110W   &05:21:31  &STEP32  &11 &13  & 95.96 &1247.5  
&42.5 \\
55~Cnc  &F204M   &05:35:49  &STEP32  &10 & 4  & 63.96 & 255.8  
&42.5 \\
55~Cnc  &F240W   &05:40:56  &STEP64  &11 & 5  &127.96 & 639.8  
&42.5 \\
\\
55~Cnc  &F110W   &06:48:20  &STEP32  &11 & 4  & 95.96 & 383.8  
&57.5 \\
55~Cnc  &F160W   &06:55:31  &STEP64  &12 & 1  &191.96 & 192.0  
&57.5 \\
\\
HD~75216 &F110W   &07:14:14  &STEP32  &11 & 4  & 95.96 & 383.8  
&57.5 \\
HD~75216 &F110W   &07:22:13  &STEP32  &10 & 5  & 63.96 & 319.8  
&57.5 \\
\enddata
\tablenotetext{a}{Number of repeated multiaccum exposures}
\tablenotetext{b}{Total Integration time for that 
filter/orientation}
\tablenotetext{c}{Angle of image +Y axis eastward of celestial 
north}
\end{deluxetable}

\clearpage
\begin{deluxetable}{lcccc}
\tablenum{3}
\tablewidth{0pc}
\tablecaption{Reference PSFs from NICMOS GTO/7226 
(n418)\label{tab_psfs}}
\tablehead{
\colhead{Star}
               &\colhead{Visit}
                      &\colhead{Spec}
                          &\colhead{H (mag)}
                                  &\colhead{Obs Date}
}
\startdata
\it HD~75216       &\it 62    &\it K2III  &\it 4.60  &\it 01 NOV 
98 \\
\it $\tau^1$ Eri   &\it 80    &\it F5V    &\it 3.35  &\it 14 NOV 
98 \\
\it $\tau^1$ Eri   &\it 81    &\it F5V    &\it 3.35  &\it 14 NOV 
98 \\
\it $\epsilon$ Eri &\it 03    &\it K2V    &\it 1.6   &\it 26 SEP 
98 \\
$\tau^1$ Eri   &18    &F5V    &3.35  &24 SEP 98 \\
$\epsilon$ Eri &04    &K2V    &1.6   &26 SEP 98 \\
UX ORI         &11/12 &A2Ve   &7.95  &04 NOV 98 \\
49 CET         &14/15 &A1V    &5.59  &07 NOV 98 \\
VEGA           &73    &A0V    &0.03  &09 NOV 98 \\
HR 4748        &92    &B8V    &5.44  &16 AUG 98 \\
\enddata
\end{deluxetable}

\clearpage
\begin{deluxetable}{lcc}
\tablenum{4}
\tablewidth{0pc}
\tablecaption{Integrated Flux Densities 
(mJy)\label{tab_flux}}
\tablehead{
\colhead{PSF}
      &\colhead{55~Cnc Visit 60}
                               &\colhead{55~Cnc Visit 61}}
\startdata
V62   &$+$0.66 $\pm$ 0.046 &$+$0.40 $\pm$ 0.058 \\
V81   &$+$0.23 $\pm$ 0.075 &$+$0.02 $\pm$ 0.074 \\
V80   &$-$1.26 $\pm$ 0.093 &$-$1.48 $\pm$ 0.084 \\
V03   &$+$0.88 $\pm$ 0.080 &$+$0.72 $\pm$ 0.077 \\
\sidehead{Model TB98 Disk\tablenotemark{a}: 8.2 mJy (F110W; 
28$<$r$<$42 AU)}
\enddata
\tablenotetext{a}{Flux(r) = 1.24 (r/28AU)$^{-5}$ mJy arcsec$^{-2}$ 
on
major axis, sinusoidally modulated azimuthally with 1.7:1 
axial fluxratio.}
\end{deluxetable}
\clearpage
\end{document}